\documentclass{IAU-TrA}
\usepackage{graphicx}

\title[Solids and Their Surfaces] %% header right hand page %%
{}

\author[DIVISION~B / COMMISSION~14 / WORKING GROUP] %% header left hand page %%
{}

\pubyear{2015}
\volume{Volume XXIXA}
\pagerange{001--008}
\jname{Transactions IAU, Volume XXIXA}
\editors{Thierry Montmerle, ed.}
\begin{document}

\maketitle

{\bf

\large
\noindent
DIVISION~B / COMMISSION~14 / WORKING GROUP                            \\ 
SOLIDS AND THEIR  SURFACES                                       \\

\normalsize

\begin{tabbing}
\hspace*{45mm}  \=                                                   \kill
CHAIR           \> Gianfranco Vidali                                  \\
VICE CHAIRS \>Harold Linnartz and Thomas Henning
\end{tabbing}

\vspace{3mm}

\noindent
TRIENNIAL REPORT 2012-2015
}

\firstsection % if your document starts with a section,
\section{Introduction}

Just a decade ago, only a handful of laboratories were engaged in the preparation and characterization of cosmic dust analogues and in the study of the formation of molecules on their surfaces. Now more than three dozen  laboratories (see list below) work on such topics. 

If in the past attention was primarily devoted to obtain the efficiency of molecule formation (as in the case of molecular hydrogen formation on amorphous silicate surfaces, Katz et al. ApJ 522, 305 (1999) or in exploring which molecules were produced in the irradiation of cosmic ices with UV or cosmic ray analogues, Allamandola et al., Space Sci. Rev. 90, 219 (1999); Moore \& Hudson, Proc. IAU Symposium 231, 247 (2006)), now it has become clear that, in oder to study molecules formation on dust analogues, it is necessary to learn about atomistic processes of adsorption, diffusion, reaction and desorption from surfaces of such solids (G.Vidali, Chem.Rev.113, 8762 (2013)). As an illustration of the importance of such knowledge in astrochemistry, we cite the process of diffusion. Diffusion of radicals on surfaces and in solids (such as in cosmic ices) is of the uttermost importance in the formation of molecules on dust grains. Hama and Watanabe reviewed (Chem.Rev. 113, 8783 (2013))  how to obtain by appropriately designed experiments activation energy barriers for the motion of hydrogen on the surface of amorphous ice,  while at the recent Faraday Discussion 168 {\em Dust, ice and gas}(see below), He \&  Vidali (Faraday Discuss. 168, 517 (2014)) presented a model on how to tease out information on diffusion from laboratory data.

\subsection{Cosmic dust}
Cosmic dust particles are providing the first solid surfaces in space and allow the formation of molecular hydrogen and complex organic molecules in a variety of environments, ranging from dense and shielded molecular clouds to protoplanetary disks and circumstellar shells around evolved stars. Thermal dust radiation from near-infrared to millimeter wavelengths as well as infrared extinction mapping allow to determine column densities, masses, and temperatures of dusty object from high-z quasars to the most nearby debris disks.

In the period of this report, the Herschel Space Observatory provided impressive dust maps of molecular clouds in our and other galaxies and ALMA is starting to map the dust emission of protoplanetary disks in unprecedented detail. The Herschel continuum data allowed to build a very comprehensive dust model for the Andromeda galaxy (Draine et al.: ApJ 780, 172, 2014) and allowed  to constrain the temperature and density profile of dense molecular cloud cores such as B68 (Nielbock et al. A\&A 538, 11, 2012), just to give two examples. The Planck collaboration delivered an all-sky model of thermal dust emission (A\&A 571, A11, 2014).

"Coreshine" - scattered light measured at mid-infrared wavelengths with Spitzer - remains an important tool to characterize the size distribution of dust grains in molecular cloud cores and provides strong evidence for the presence of micron-sized grains in these environments (e.g. Lefevre et al.: A\&A 572, A20, 2014;  Steinacker et al. 2015, submitted). 
Although Herschel operated at a wavelength outside the wavelength range where most of the solid-state dust features occur - this range  was well covered by the Spitzer satellite - spectroscopy of the 69 micron forsterite feature was possible and demonstrated that the crystalline silicates  in protoplanetary disks are nearly iron-free (Sturm et al.: A\&A 553, A5, 2013 ). In addition, evidence for crystalline water ice in disks  has been found through the detection of its far-infrared band (e.g. Min et al. 2014, submitted). 
The discussion of the far-infrared and submillimeter opacities and their temperature dependence is continuing and their behavior remains a largely unsolved problem, although new laboratory data have been provided.

Major progress has been reached in our understanding of the formation process of cosmic dust. It now becomes clear that not only at high redshift, but also in the Milky Way, the SMC and LMC, AGB stars fall short to explain the cosmic budget (e.g. Zhukovska: A\&A 562, A76, 2014 Zhukovska \& Henning: A\&A 555, 99, 2013). Herschel observations clearly demonstrated that core-collapse supernovae are able to produce considerable amounts of dust, although its survival during the reverse shock phase remains far from obvious. A particularly impressive example is the measurement of SN 1987A where about 0.4-0.7 solar masses of cold dust were detected (Matsuura et al.: Science 333, 1258,  2011; Matsuura et al. 2014, in press) and could be located inside the HST ring of shocked emission with ALMA observations (Indebetouw et al.: ApJ 782, L2, 2014 2014: $>$0.2 M$_{sun}$). It remains still an open question if dust is predominantly formed in core-core collapse supernovae or the general interstellar medium. Laboratory experiments indicate that dust formation at low temperatures is a possible process (Krasnokutski et al.: ApJ 782, 15, 2014).  
The most important event for the astrophysical dust community was the conference "{\em The Lifecycle of Dust in the Universe: Observations, Theory, and Laboratory Experiments}", held in Taipei (Taiwan) between November 18th and 22nd, 2013. The meeting was organized five years after the last big dust conference in Heidelberg and brought together more than 150 participants in the field of observational astronomy, theoretical astrophysics, meteoritics, and laboratory astrophysics, The meeting was extremely productive and should the increased interest of the extragalactic community in dust-related research. The diffuse interstellar bands - hundreds of spectroscopic features at visible wavelengths - remain a puzzle, although an entire IAU symposium (297) has devoted to this topic. The meeting took place in the Netherlands between May 20th and May 24th, 2013 with a strong participation of observers and experimentalists.

\subsection{Cosmic ice}
Cold dust particles act as cryosurfaces onto which molecules accrete, meet and greet, i.e., freeze out, diffuse and react.  Astrochemists explain the observed chemical complexity in space - so far 180 different species have been identified - as the cumulative outcome of reactions in the gas phase and on icy dust grains. Gas phase models explain the observed abundances of a substantial part of the observed species, but fail to explain the observed abundances of stable species, as simple as molecular hydrogen and water, methanol or acetonitrile - a precursor molecule for the simplest amino acid glycine - as well as larger compounds such as glycolaldehyde, dimethylether and ethyleneglycol. Evidence has been found that these and other complex, organic compounds form on icy dust grains  upon (non)energetic processing, such as irradiation by vacuum UV light, interaction with particles (atoms, electrons and cosmic rays) and and heating. (E.F. van Dishoeck, Faraday Disc. 168, 9 (2014). The surface acts as a molecule reservoir and a catalytic sites for molecule formation. 
Over the last years more and more laboratories worldwide have been performing experiments to characterize the solid state processes at play (Linnartz, Ioppolo, and Fedoseev, Int. Rev. Phys. Chem., submitted.) . Such work not only provides fundamental insights in the underlying physical and chemical processes, but also allows to interpret and steer astronomical observations. New flagship facilities, such as ALMA, unravel the molecular complexity of the inter- and circumstellar medium that finds its origin also in the solid state. The detection of 1000s of exo-planets over recent years only confirms that a decent understanding of processes on (icy) dust grains is highly needed to understand how exo-planetary atmospheres form. Physically, ice covered particles stick easier together, chemically, these provide the material from which planets ultimately form.

In 2014 the Faraday Discussions 168 on {\em Dust, ice and gas} took place in the Netherlands, summarizing the state-of-the-art in laboratory based dust and ice research. The resulting Faraday Issue (rsc.li/fd168) gives a good view of recent accomplishments in the study of dust and interactions of photons, atoms, free radicals, molecules, cosmic rays or electrons with astrophysical surfaces. Since 2000 an impressive increase of scientific achievements has been realized in this field, largely stimulated also by the introduction of highly sophisticated surface techniques: Thermal Programmed Desorption (TPD), Reflection Absorption Infrared Spectrometry (RAIRS), Resonant Enhanced Multiphoton Ionization (REMPI), time-of-flight desorption spectroscopy and Atom Force Microscopy (AFM), to mention a few (Laboratory Astrochemistry, Wileys, 2014, Eds. Schlemmer, Mutschke, Giesen). Also the use of large scale research infrastructures, such as synchrotrons and free electron lasers has entered the field (Fillion et al. Faraday. Disc. 168, 533 (2014) and H.M. Cuppen et al. Chem. Rev. 113, 8840 (2013).). These techniques yield information about the kinetics and energetics of atomic/molecular diffusion on and desorption from surfaces, reaction products, the ro-vibrational state of ejected products, and the morphology of the solid surfaces. Publications are not only available from the astronomical literature but also in the physical chemical and surface science journals. Moreover, fundamental solid state processes are nowadays also included in the large astrochemical databases, linking networks historically dominated by gas phase processes to reactions in the solid state (Wakelam et al. Space Science Rev. 156, 13 (2010)). Also in theory, new developments are found. First of all, basic solid state physics and chemistry Ð e.g., thermal and photodesorption or hydrogenation processes - are simulated at the level of individual atoms and molecules for conditions that are typical for space. New stochastic tools predict molecule formation processes Ð taking into account different reaction mechanisms (Langmuir-Hinshelwood, Ealy-Rideal, and hot-atom) Ð capable of extending laboratory findings to timescales as needed to explain astronomical observations (H.M. Cuppen et al. Chem. Rev. 113, 8840 (2013)).

\subsection{The Laboratory - Astronomy Connection}
The measurement of the efficiency of a reaction or the characterization of a cosmic dust analogue need eventually to be related to astronomical observations. This is accomplished via simulations that take appropriately analyzed data from the laboratory and use them to predict the chemical make-up and evolution of interstellar medium environments. As molecules can be produced in a variety of ways on/in ices using, for example, neutral reactants or via the interaction of UV/charged particle radiation with ices, it is important to have simulations that compare quantitatively the different competing mechanisms. In a recent laboratory study of the formation of hydroxylamine via oxidation of ammonia ice (He et al., ApJ 799, 49 (2105)), a simulation was carried out that included both the formation of NH$_2$OH via  NH$_3$ oxidation (He et al., ApJ 799, 49 (2015)) and via sequential hydrogenation of NO ice (Congiu et al., J.Chem.Phys. 137, 054713 (2012)). Such comparisons are not done frequently enough at this detailed level, but, as more data are coming out of laboratories, more of these types of studies will be necessary.

\subsection{Meetings, books, special issues and reviews}
The number of groups that have become active in the dust and ice field has been steadily increasing over the last years and an overview at this place Ð like in previous reports Ð is no longer an option anymore. Many new initiatives have been made. A large European network Ð LASSIE, Laboratory Astrophysics Surface Science in Europe Ð training 30 new PhD students in the field of astronomical dust and ice, came to a successful end in February 2014, and meanwhile other initiatives, among others at NASA (Carbon in the Universe) have started. Interstellar dust and ice is clearly a hot topic. Other large meetings, besides the IAU297, FD168 and 2013 Taipei meeting, have been the ICE2013 and ICE2015 workshops in Kauai'i, the bi-annual solid theme sessions within the American Chemical Society  Astrochemistry sub-division meetings, and the yearly Laboratory Astrophysics Division meeting of the American Astronomical Society. A large number of special  issues, books and reviews have been devoted to dust and ice in space, generally within the context of astrochemistry, but with clear focus chapters on solid state processes. For example, in 2014 ÔLaboratory Astrochemistry Ð from molecules through nanoparticles to grainsÕ has been published. In 2013, Chemical Reviews dedicated a full issue to ÔAstrochemistryÕ with several reviews on the solid universe. Also in that year, Õ45 years of astrochemistryÕ, a special issue in honor of Prof. Oka appeared (J.Phys.Chem. A 117, 9308 (2013)). The many references listed in all these publications show that the dedicated efforts to link observational, laboratory and modeling studies to characterize the physics and chemistry of dust and ice in space is a very active research area, substantially contributing to our understanding of the universe.  

\section{List of laboratories} 
The following is a list of laboratories active in 2012-2015 and working on the interaction of particles (charged and neutral particles, photons) with surfaces of dust grain/ice analogs and on the formation and characterization of dust/ice analogues. The main research area of each lab is given in parenthesis (MF=molecule formation on dust grain and ice analogues; DF=dust formation and dust and ice characterization). For more information, visit the groups' Web pages.
\begin{itemize}
\item Syracuse University, USA, G.Vidali (MF)
\item	Hokkaido University, Japan, N. Watanabe / A. Kouchi (MF)
\item	NASA Ames Research Center, USA, L. Allamandola, F. Salama (MF)
\item	NASA - Goddard Space Flight Center,  USA, R. Hudson (MF)
\item	University of Hawai'i, USA, R. Kaiser (MF)
\item	Leiden University, The Netherlands, H. Linnartz (MF)
\item	Heriot-Watt University, UK, M. R. S. McCoustra (MF)
\item	University of Chergy-Pontoise, France, J.L. Lemaire (MF) 
\item	Aarhus University, Denmark, Interdisciplinary Nanoscience Center (INANO-Fysik),  L. Hornekaer (MF)
\item	Aix-Marseille University, France, Laboratoire des Physique des Interactions Ioniques et MolŽculaires, France, T. Chiavassa (MF)
\item	Centre Universitaire dÕOrsay Institut d'Astrophysique Spatiale (IAS), France, L. d'Hendecourt (MF)
\item	Centro de Astrobiolog'a (INTA-CSIC), Spain, Madrid, G. M. Munoz-Caro (MF)
\item	Instituto de Estructura de la materia CSIC, Group of Molecular Physics of Atmospheres and Plasmas, Madrid, Spain, R. Escribano (DF)
\item	National Institute for Astrophysics (INAF), Catania Astrophysical Observatory, Italy, M. E. Palumbo (MF)
\item INAF-Osservatorio Astronomico di Capodimonte, Italy V. Mennella (MF)
\item University of Virginia, USA, R. Baragiola (MF)
\item	Open University, Centre for Molecular and Optical Sciences (CeMOS), UK, N. J. Mason (MF)
\item	Paris Observatory, France, J. L. Lemaire (MF)
\item	University of M$\ddot{u}$nster, Physikalisches Institut, Germany, H. Zacharias (MF)
\item	University of Sussex, UK, W. Brown (MF)
%\item	Radboud University Nijmegen, Theoretical Chemistry, Institute of Molecules and Materials, The Netherlands, H. M. Cuppen 
\item	UniversitŽ Paris-Sud, Institut d'Astrophysique Spatiale (IAS), France,  E. Dartois (DF)
\item	UniversitŽ Pierre et Marie Curie, Laboratoire de Physique MolŽculaire pour lÕAtmosphre et lÕAstrophysique (LPMAA), France, J. H. Fillion (MF)
\item	University College London, Department of Chemistry, UK, S. D. Price (MF)
\item University of Jena, Laboratory Astrophysics, Germany, H. Mutschke (DF)
\item	Max Planck Institute for Astronomy, Laboratory Astrophysics and Cluster
Physics Group at the University
of Jena, Germany, T. Henning \& C. Jaeger (DF)
\item University of Missouri, Columbia, USA, A. Speck (DF)
\end{itemize}

\section{Meetings}

\vskip 0.1 true in
During this triennial, the American Astronomical Society (USA) approved the constitution of the Laboratory Astrophysics Division. LAD organizes meetings within the AAS Summer National Meetings. The American Chemical Society (USA) approved the formation of a subdivision of the Physical Chemistry Division, the Astrochemistry sub-division. It holds regular meetings during the semi-annual  meetings of ACS. 
\vskip 0.2 true in
List of important meetings dedicated to the interaction of particles with solids (listed in inverse chronological order):
\begin{itemize}
\item "Second  Experimental Laboratory Astrophysics Workshop - ICE2015", Kauai'i, USA, 2015
\item "Workshop on Interstellar Matter", Hokkaido University, Japan, 2014
\item COSPAR2014, Moscow, Russia, 2014
\item American Chemical Society, Symposia at National Meetings, USA, 2014
\item Faraday Discussion on "Dust, Ice and Gas", the Netherlands, 2014
\item "The Lifecycle of Dust in the Universe", Taipei, Taiwan, 2013
\item American Astronomical Society, Laboratory Astrophysics Division Meetings, USA, 2012, 2013, 2014, 2015
\item "Dust Growth in Star- \& Planet-Forming Environments 2013", Heidelberg, Germany 2013
\item "Cosmic Dust", Kobe, Japan 2013
\item "First  Experimental Laboratory Astrophysics Workshop -ICE2013", Kauai'i, USA, 2013
\item "The Warm Universe: Astrochemistry at Intermediate and Elevated Temperatures", Tallinn, Estonia, 2012
\item COSPAR2012, Mysore, India, 2012
\item "Cosmic Dust", Kobe, Japan, 2012
\item IAU Symposium 292: Molecular Gas, Dust, and Star Formation in Galaxies, Bejing, China, 2012

\end{itemize}

\section{Notable publications}
\vskip 0.1 true in
A complete bibliography of works on dust and its interaction with particles is beyond the scope of this report. Several references have been given in the text and here we list  a few key review articles that appeared in the literature recently and that can be used as starting points for a deeper search.
\begin{itemize}
\item {\em Astrochemistry}, themed issue is  Phys. Chem. Chem. Phys., vol. 16 (2014)
\item {\em Astrochemistry of dust, gas and ice},  themed issue, Faraday Discuss., vol. 168 (2014)
\item {\em Astrochemistry}, themed issue, Chem. Rev., vol. 113 (2013)
\item A.C. A. Boogert, P. A. Gerakines, and D. C.B. Whittet
Annu. Rev. of Astron. \& Astrophys. Vol. 53 (2015)
\item B-G Andersson, A. Lazarian, and J. E. Vaillancourt, Annu. Rev. of Astron. \& Astrophys. Vol. 53 (2015) 
\item G.Vidali  J. Low Temp. Phys., Vol. 170, 1 (2013)
\item A. G. G. M. Tielens, Rev. Mod. Phys. Vol. 85, 1021  (2013)

\end{itemize}
 \vspace{6mm}
 
{\hfill Gianfranco Vidali}

{\hfill {\it Chair of Working Group}}

{ \hfill {\it Solids and Their Surfaces}}

{\hfill Harold Linnartz and Thomas Henning}

{\hfill {\it Vice-Chair of Working Group}}

{ \hfill {\it Solids and Their Surfaces}}
\end{document}